%% file: article.tex
\documentclass[journal]{IEEEtran}

\usepackage[numbers,sort&compress]{natbib}
\usepackage{amsfonts}
\usepackage{multirow}

\ifCLASSINFOpdf
	\usepackage[pdftex]{graphicx}
\else
\fi

\begin{document}
	
\title{Pyramid Multi-branch Fusion DCNN with Multi-Head Self-Attention for Mandarin Speech Recognition}
	
\author{
	\IEEEauthorblockN{
		Kai Liu,
	 	Hailiang Xiong,
		Gangqiang Yang,
		Zhengfeng Du,
		Yewen Cao,
		Danyal Shah
	}
\\
}

\maketitle
\begin{abstract}
As one of the major branches of automatic speech recognition, attention-based models greatly improves the feature representation ability of the model. In particular, the multi-head mechanism is employed in the attention, hoping to learn speech features of more aspects in different attention subspaces. For speech recognition of complex languages, on the one hand, a small head size will lead to an obvious shortage of learnable aspects. On the other hand, we need to reduce the dimension of each subspace to keep the size of the overall feature space unchanged when we increase the number of heads, which will significantly weaken the ability to represent the feature of each subspace. Therefore, this paper explores how to use a small attention subspace to represent complete speech features while ensuring many heads. In this work we propose a novel neural network architecture, namely, pyramid multi-branch fusion DCNN with multi-head self-attention. The proposed architecture is inspired by Dilated Convolution Neural Networks (DCNN), it uses multiple branches with DCNN to extract the feature of the input speech under different receptive fields. To reduce the number of parameters, every two branches are merged until all the branches are merged into one. Thus, its shape is like a pyramid rotated 90 degrees. We demonstrate that on Aishell-1, a widely used Mandarin speech dataset, our model achieves a character error rate (CER) of 6.45\% on the test sets.
\end{abstract}
\IEEEpeerreviewmaketitle
 \section{Introduction}
 \IEEEPARstart{M}{andarin} automatic speech recognition (ASR) has developed with the popularity of deep learning, especially in recent years, new architectures have been proposed, including DeepSpeech\cite{2014deepspeech}, DeepSpeech2\cite{2015deepspeech2}, WeNet\cite{2021WeNet}, etc. Compared with Convolutional Neural Networks (CNN) \cite{2012ApplyingCNN, 2013DeepLVCSR, 2014ConvolutionalNN, 2016Towards, 2019LiJasper, 2019QuartzNet}, people usually use Recurrent Neural Networks (RNN)\cite{2017RaoExploring, 2019HeStreaming, 2020SainathStreaming} for sequence-to-sequence tasks. The reason for this difference is that the RNN assumes that the development of things unfolds in a time series, that is, what happens at a previous moment will have an impact on the development of things in the future. CNN, on the other hand, assumes that human vision will always focus on the most obvious point in the line of sight. Therefore, RNN is easy to obtain a higher recognition rate in speech recognition. But after the transformer\cite{2017Attention} was proposed, the widespread application of the attention mechanism broke people's previous ideas, especially in the Conformer\cite{2020Conformer}, the combination of CNN and attention mechanism achieved the state-of-the-art model at the time, which made CNN gain great attention in the field of speech recognition again.

\begin{figure*}[!t]
	\centering
	\includegraphics[width=\linewidth]{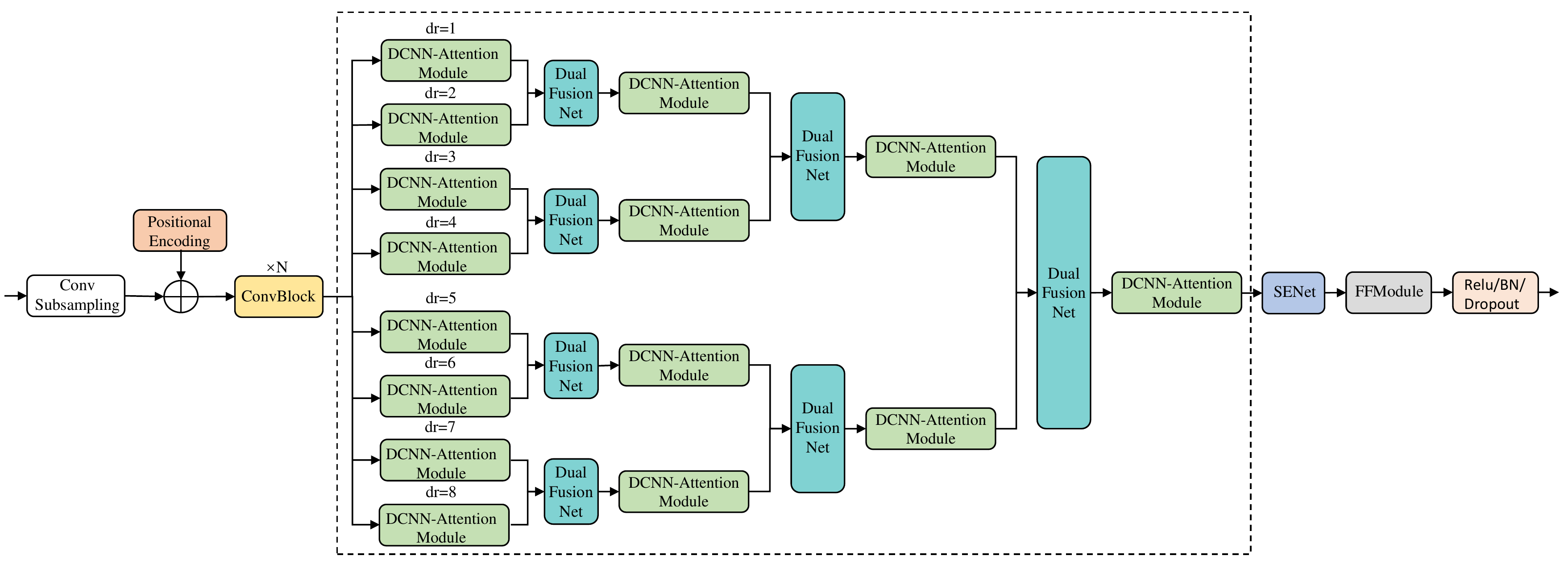}
	\label{fig_1}
	\hfil
	\caption{Pyramid multi-branch fusion DCNN with multi-head self-attention. The figure shows that the number of branches is 8, Then every two branches are combined until they merge into one branch, and the dilated rate of the eight branches in the first layer is [1, 2, 3, 4, 5, 6, 7, 8], the dilated rate of the four DCNN in the second layer is [1, 2, 3, 4], etc. }
\end{figure*}
CNN has excellent characteristics of extracting local features, exploiting the invariance of convolution to overcome the diversity of the speech signal itself. CNN have been used for speech recognition for a long time, but only as part of a traditional pipeline. They were first introduced by Time-Delay Neural Networks (TDNNs) to predict phoneme classes\cite{1990TDNN}, and later to generate hidden Markov model (HMM) posteriorgrams. CLDNN\cite{2015CLDNN} is a typical application of early CNN in speech recognition, it uses CNN to extract short-term features of input speech and combines Long Short Term Memory RNN(LSTM)\cite{1997LSTM} to achieve better performance than LSTM in speech recognition. For Mandarin speech recognition, DFCNN\cite{2018dfcnn} firstly applied CNN to large-scale continuous speech recognition(LVCSR), it uses a large number of CNNs to directly model the whole sentence speech signal, which better expresses the long-term correlation of speech and achieves excellent recognition results. However, a model with CNN has its limitations. One limitation of using local connectivity is that a large number of layers or parameters are needed to capture global information. To combat this issue, contemporary work ContextNet\cite{2020ContextNet} adopts the squeeze-and-excitation module (SENet)\cite{2018SENET} in each residual block to capture longer context. However, it is still limited in capturing dynamic global context as it only applies a global averaging over the entire sequence.

With the introduction of the attention mechanism, this defect of CNN has been considerably compensated. And its characteristics can enable the model to obtain the global context information of the input features well. Some recent studies have demonstrated that the combination of CNN and attention mechanism performs better than using them individually \cite{2020Augmented}. In \cite{2020LiteTransformer}, a multi-branch architecture was proposed which splits the input into two branches: the self-attention module and convolution module, and the output of two modules are concatenate to obtain the prediction results. Not only that, there is a combination of Connectionist Temporal Classification (CTC)\cite{2006CTC} and attention to improve its performance. In \cite{2017Hybrid, 2016CTC-Attention}, the hybrid CTC/attention architecture was proposed to leverage the CTC objective as an auxiliary task in an attention-based encoder-decoder network. In \cite{2018StateSpeech}, the well-known Listen, Attend and Spell (LAS) ASR model \cite{2016LAS} applied the multi-head approach to the attention mechanism to further improve its already state-of-the-art accuracy on the large-scale voice search data. But multi-head self-attention has its limitations, for speech recognition of complex languages, on the one hand, a small head size will lead to an obvious shortage of learnable aspects; on the other hand, we need to reduce the dimension of each subspace to keep the size of the overall feature space unchanged when we increase the number of head, which will greatly weaken the ability to represent the feature of each subspace.

In response to the above flaws, we add a DCNN\cite{2016DCNN} before each multi-head self-attention. Multiple branches in parallel are also applied to the model, each branch has a different dilated rate. For each branch, the multi-head self-attention will obtain the feature information of completely different receptive fields, forcing the multi-head self-attention to learn semantics at different aspects of information. That is, the multi-head self-attention on each branch does not need to extract global feature information, but only needs to extract the global feature of a certain semantic aspects within the receptive field of the current branch. At the same time, to prevent a sharp increase in the number of parameters caused by multiple parallel branches, every two branches are merged until all the branches are merged into one. The decoders we adopt are CTC, which can automatically align the labels with the model's output. We tested on Aishell-1\cite{2017AISHELL-1}, a widely used Mandarin speech dataset. Our model achieves a character error rate (CER) of 6.45\% on the test sets.

The contribution of this paper is three-fold:
\begin{itemize}
	\item First, we propose pyramid multi-branch fusion DCNN with multi-head self-attention. It is a new structure combining DCNN and multi-head self-attention. This structure can enable multi-head self-attention to obtain feature information in different receptive fields. To reduce the number of model parameters, a branch fusion strategy is used, every two branches are merged into a new branch, and the last layer is merged into a branch output.
	\item Second, we conduct comprehensive experiments to verify our design, including verifying that the DCNN we used works properly, selecting the most appropriate dilated rate, etc. Experiments results show that our model is capable of effective speech recognition.
	\item Third, experiments and analyses are carried out on the benchmark Chinese speech corpora Aishell-1. Compared with well-practised ASR models, the proposed model not only achieves competitive or superior results but also demonstrates enhanced efficiency and effectiveness.
\end{itemize}

This paper is organized as follows. We begin with a review and analysis of related works in ASR, especially Mandarin ASR in Section 2. In Section 3, we describe the details of our pyramid multi-branch fusion DCNN with a multi-head self-attention system focusing on the neural network architecture. In Section 4, we provide the experimental setup and discuss the results from various configurations in the proposed method. In Section 5, we conclude the paper with summary remarks and future directions.

\begin{figure*}[]
	\centering
	\includegraphics[width=16cm]{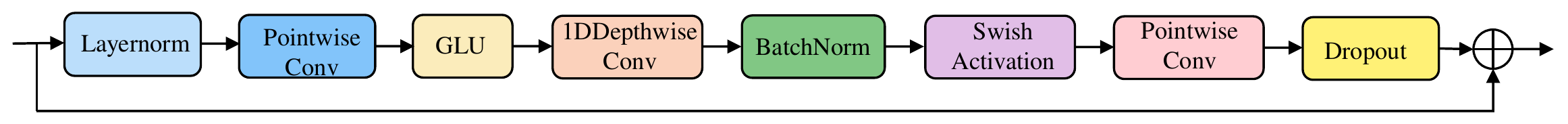}
	\label{fig_2}
	\hfil
	\caption{ConvBlock. This module consists of: Layernorm, Pointwise convolution, GLU, Depthwise convolution, BatchNorm, Swish activation function, and Dropout, where the default value of the Depthwise convolution expansion factor is 2.}
\end{figure*}


\section{Pyramid Multi-branch Fusion DCNN with Multi-Head Self-Attention}

Our proposed model, named \textit{Pyramid Multi-branch Fusion DCNN with Multi-Head Self-Attention}, is mainly divided into three parts, its structure is shown in Fig. 1. Before the multi-branch network, the shallow speech features are first extracted by the convolution module. This module is similar to the Convolution Module in the Conformer\cite{2020Conformer}. Then, after DCNN with different dilated rate across multiple streams, the obtained speech features of different temporal resolutions are then used to extract global features by multi-head self-attention, they merge into one layer in the end. After Multi-branch, SENet is used to receive its output, this network is equivalent to the channel attention mechanism, which can combine the weights between each channel of all features.

\subsection{\textbf{Before Multi-branch}}

This part mainly consists of two modules: PositionalEncoding, and ConvBlock. We adopted the positional encoding in \cite{2017Attention}, which is an extra step of CNN compared with RNN. Due to the sequential tasks, the features are input by the model over time, this process has temporal information, and CNN cannot obtain this information, we have to add this part of the information using position encoding manually, its formula is:
\begin{equation}
	P(in_{pos}, 2i) = sin(in_{pos}/10000^{2i/d_{model}})
\end{equation}
\begin{equation}
	P(in_{pos}, 2i+1) = cos(in_{pos}/10000^{2i/d_{model}})
\end{equation}
\noindent
where $in_{pos}$ is the position, and $i$ is the dimension. Each dimension of the positional encoding corresponds to a sinusoid. The wavelengths form a geometric progression from $2\pi$ to $10000 \cdot 2 \pi$. For any fixed offset $K$, $P_{in_{pos}}+K$ can be expressed as a linear function of $P_{in_{pos}}$.

We use ConvBlock to initially extract shallow features. ConvBlock is the module proposed in Conformer ASR model\cite{2020Conformer}, Fig. 2 illustrates the ConvBlock. The convolution module contains a pointwise convolution with an expansion factor, we try the combination of different expansion factors in the multi-layer stacked ConvBlock to obtain better recognition effect, the results are shown in Table \uppercase\expandafter{\romannumeral7}. And then GLU activation layer projects the number of channels, followed by a 1D depthwise convolution. The 1D depthwise convolution is followed by a batchnorm and then a swish activation layer. Let the input sequence be $\textbf{x} = (x_1,...,x_T)$, the ConvBlock transforms the original signal $\textbf{x}$ into a high level representation $\textbf{h} = (h_1,...,h_{T^{`}})$, where $T^{`} \leq T$. Its formula is:
\begin{equation}
	\textbf{h} =  CB_i(CB_{i-1}(...CB_1(\textbf{x})))
\end{equation}
where each $CB_i(\cdot)$ defines a convolution block, $i$ represents the number of stacked ConvBlock.

\subsection{\textbf{Multi-branch Fusion DCNN with Multi-Head Self-Attention}}

In order to enable our model to extract semantic expressions of different dimensions from the input features as much as possible and to avoid the problem of insufficient representation dimension of the attention subspace when the number of heads is set too large for multi-head self-attention. We use a multi-branch network to enable self-attention to extract only the speech features of a certain branch, which significantly reduces the number of features it needs to extract. This allows us to fully characterize the semantic information of the current branch by only setting a small subspace dimension. To explicitly distinguish the difference of each branch, we deploy DCNN to ensure that each branch has a distinct receptive field. Fig. 3 shows an example of a 1D dilated convolution with a $3 \times 1$ kernels and a dilated rate of 2.

\begin{figure}[h]
	\centering
	\includegraphics[width=8.5cm]{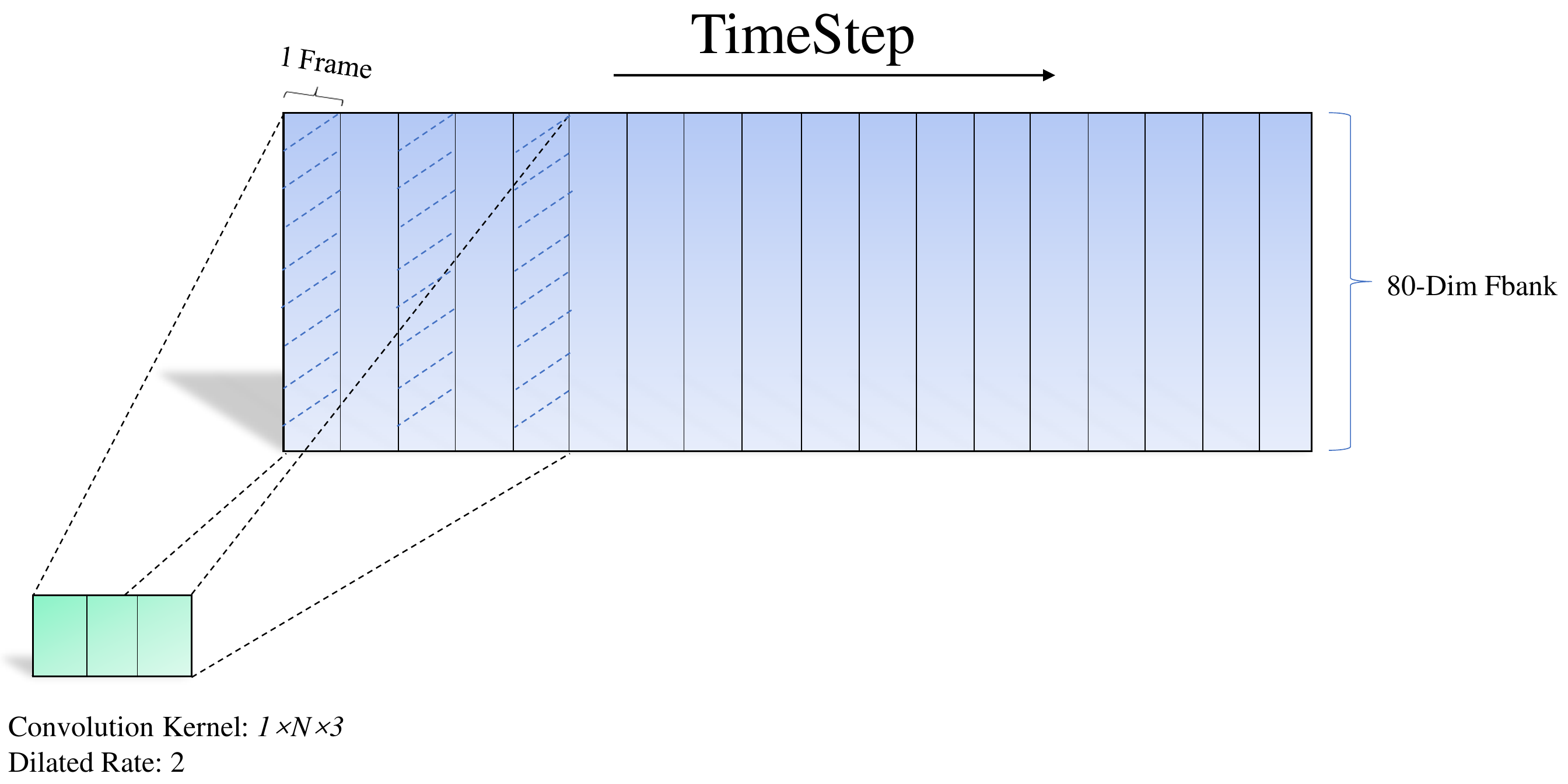}
	\label{fig_3}
	\hfil
	\caption{1D Dilated Convolution. The figure shows that the convolution kernel size is 3 and the dilated rate is 2.}
\end{figure}

Unlike standard convolution kernels, DCNN adds some holes in the kernel, which can expand the receptive field of the model, DCNN can be expressed as:
\begin{equation}
	F^{'}(t) = (F*k_{l})(t) = \sum_{n=-r}^{r}F(n+t)k_{l}(n)
\end{equation}
where $F^{'}(t)$ is the eigenvalue of the feature map at $t$ after dilated CNN. $F(t)$ denotes the eigenvalue representation of the feature map at $t$ without dilated CNN. $k_l$ represents a 1D convolution kernel with a dilated rate of $l$, and the size of the convolution kernel is $(2n+1)\times 1$, followed by a MHSAModule, which is used in Conformer. This module utilizes the scaled dot-product attention to map the input sequence $ \textbf{X}\in \mathbb{R}^{N \times d_{model}}$, where $N$ is the total number of input embeddings restricted by the left and right context and $d_{model}$ is the dimension of embeddings used inside the self-attention. For three inputs queries $\textbf{Q}_i$, keys $\textbf{K}_i$, and values $\textbf{V}_i$ with dimension $d_q$, $d_k$ and $d_v$, the outputs for the $i$-th head can be defined as:
\begin{equation}
	Head_i = Softmax(\frac{\textbf{Q}_{i}\textbf{K}_i^{T}}{\sqrt{d_k}})\textbf{V}_i
\end{equation}
where $\textbf{Q}_i = \textbf{X}\textbf{W}_{i}^{Q}$ and $\textbf{W}_{i}^{Q}\in \mathbb{R}^{d_{model} \times d_{q}}$, $\textbf{K}_i = \textbf{X}\textbf{W}_{i}^{K}$ and $\textbf{W}_{i}^{K}\in \mathbb{R}^{d_{model} \times d_{k}}$, $\textbf{V}_i = \textbf{X}\textbf{W}_{i}^{V}$ and $\textbf{W}_{i}^{V}\in \mathbb{R}^{d_{model} \times d_{v}}$.
The multi-head outputs are concatenated and linearly projected, and then layer normalization\cite{2016LayerNorm} is applied to the projected embedding that is skip-connected with the input:
\begin{equation}
	\textit{Con}(\textbf{X}) = Concat(Head_1,...,Head_i)\textbf{W}^{O}
\end{equation}
\begin{equation}
	\textit{MHSA}(\textbf{X}) = Dropout(\textit{MHSA}(Layernorm(\textbf{X}))) + \textbf{X}
\end{equation}
where $\textbf{W}^{O} \in \mathbb{R}^{d_{model} \times (i\times d_{v})}$, and then we use dropout, which helps train and regularize deeper models. We call this architecture as \textit{DCNN-Attention} module; it is shown in Fig. 4.

\begin{figure}[h]
	\centering
	\includegraphics[width=6.3cm]{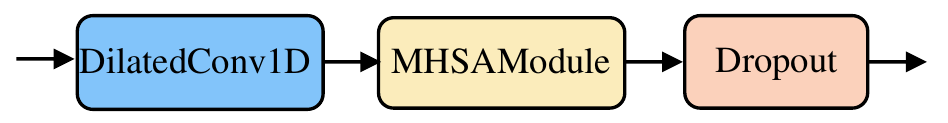}
	\label{fig_4}
	\hfil
	\caption{DCNN-Attention module. We use dilated CNN with different dilated rate to obtain receptive fields of different sizes, and use a multi-head self-attention to obtain global features.}
\end{figure}

\begin{figure}[!t]
	\centering
	\includegraphics[width=8.5cm]{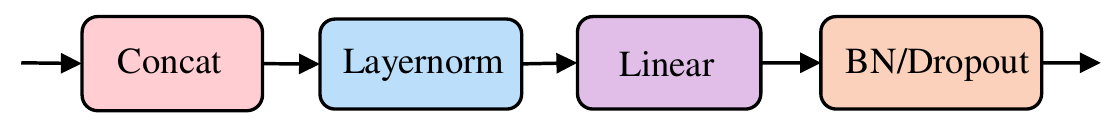}
	\label{fig_5}
	\hfil
	\caption{DualFusionNet. The two branches of the module are spliced together.}
\end{figure}
Each branch is equipped with a DCNN-attention module. Since the dilated rate of the DCNN in each module is different, resulting in various receptive fields in each branch, the semantic information that self-attention in each branch is good at extracting is also different. For example, in a large receptive field, multi-head self-attention can better extract grammatical information that is far away in time steps, such as inverted sentences. Thus, for each input, the self-attention in the branch only needs to focus on part of the semantic features, which greatly reduces the number of features. In this way, even a small dimensional vector can adequately characterize the semantic information of the current branch, that is to say, we can set a larger number of heads. Of course, the cost of adopting multiple branches is the increase in parameters, which is not in line with our original intention. Therefore, we propose a pyramidal progressive multi-branch fusion mechanism. Every two branches use a fusion module named \textit{DualFusionNet} for fusion, the module structure is shown in Fig. 5. It uses the Concatenate function to merge the two branches, followed by Layer Normalization, a linear layer for mapping to increase its linear representation capability, BatchNormalization and Dropout. For multiple parallel branches, we set it as one layer, then the output of the DCNN-attention of the $i$-th layer and the $j$-th branch is set to $O_i^{2 \times j}$, where $i = 1, 2,..., n$, $j = 1, 2,..., m/2$, then the output formula of DualFusionNet is:

\begin{equation}
	Con_{i+1}^{j} = Concat(O_i^{2 \times j-1},O_i^{2 \times j})
\end{equation}
\begin{equation}
	DFN_{i+1}^{j} = Dropout(BN(Linear(LN(Con_{i+1}^{j}))))
\end{equation}
where $n$ is the maximum number of layers in the network, $m$ is the maximum number of branches, $Con_{i+1}^{j}$ denotes the Concat output of the $(i+1)$-th layer and the $j$-th branch, $DFN_{i+1}^{j}$ denotes the DualFusionNet output of the $(i+1)$-th layer and the $j$-th branch. When we disable the progressive fusion network, the number of DCNN-attention is $n\times m$, on the contrary, the number is only $2^n-1$ when we use it. By adopting this structure, the number of the parameters will greatly decline. There is only one DCNN-attention module in the last layer, so fusion is no longer necessary. Here is a detail, we set the size of the convolution in the last DCNN-attention module to $2 \times d_{model}$. We hope to get more features so that the SENet has more representations.

\subsection{\textbf{After Multi-branch}}

The last part is mainly composed of SENet\cite{2020ContextNet}, Feed Forward module. Unlike self-attention, SENet weights the channel dimension of speech features to obtain the interdependence between its modeling channels. Fig. 6 shows its structure; it mainly includes Squeeze and Excitation. The Squeeze operation is a global average pooling, after the compression operation, the feature map is compressed into $1\times C $, where $ C $ is the number of channels, and it has a global receptive field. Excitation is similar to the gate mechanism in RNN, utilizing the weight matrix $\textbf{W}\in \mathbb{R}^{1 \times C}$ to regenerate the weights for each feature channel. To reduce the number of parameters, we set a decay factor $R$, which means to reduce the number of convolution kernels in the first FC layer to $\frac{1}{R}$. Let the input of SENet be \textbf{\textit{x}}, its formula is:

\begin{equation}
	S_{x} = \frac{1}{T}\sum_{t}x_{t}
\end{equation}
\begin{equation}
	E_{x} = Sigmoid(W_2(Swish(W_1S_x + b_1)) + b_2)
\end{equation}
\begin{equation}
	SE(\cdot) =  E_{x} \otimes \textbf{\textit{x}}
\end{equation}
where $x_t$ is the feature value of the input vector \textbf{\textit{x}} at time $t$, $S_x$ represents Squeeze, $E_x$ denotes Excitation, $\otimes$ means element-wise multiplication, $W_1$, $W_2$ are weight matrics, and $b_1$, $b_2$ are bias vectors.

\begin{figure}[h]
	\centering
	\includegraphics[width=4.5cm]{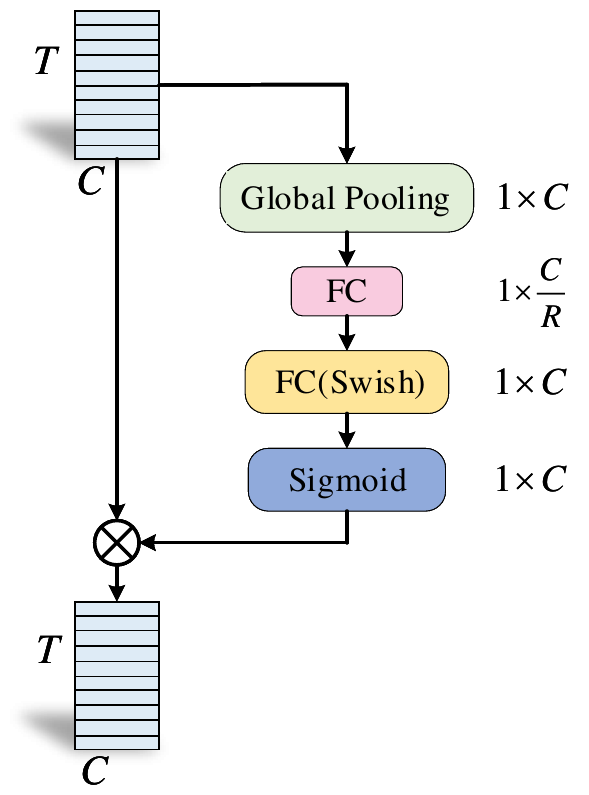}
	\label{fig_6}
	\hfil
	\caption{SENet. The Global Pooling is applied to condense the input into a 1D vector, which is then processed by a bottleneck structure formed by two fully connected (FC) layers with swish activation functions. The output goes through a sigmoid function to be mapped to (0, 1), and then applied to the input vector using pointwise multiplications.}
\end{figure}

\begin{figure*}[]
	\centering
	\includegraphics[width=12cm]{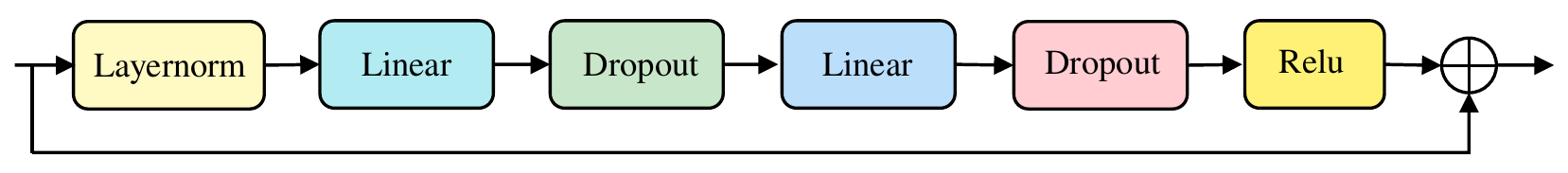}
	\label{fig_7}
	\hfil
	\caption{Feed Forward module. The first linear uses an expansion factor of 4 and the second linear layer projects it back to the model dimension, and we deploy relu activation.}
\end{figure*}

The SENet is followed by the Feed Forward module, which consists of two linear transformations and a Relu activation, as shown in Fig. 7. A residual connection is added to the feed-forward layer, followed by layer normalization. This can effectively regulate the network and prevent gradient disappearance or gradient explosion. The final Fully Connected layer (FC) is to change the feature dimension to the length of the token dictionary to apply the softmax layer. Mathematically, this means, for input $x$ of the multi-branch fusion DCNN network, the output $y$ is:
\begin{equation}
	x^{'} = CS(x) + P(x)
\end{equation}
\begin{equation}
	x^{''} = CB_{stack}(x^{'})
\end{equation}
\begin{equation}
	(x^{'''})_1^{2\times j-1} = DA(x^{''})
\end{equation}
\begin{equation}
	(x^{''''})_{i+1}^{j} = DFN_{i+1}^{j}((x^{'''})_i^{2 \times j-1},(x^{'''})_i^{2 \times j})
\end{equation}
\begin{equation}
	x^{'''''} = SE((x^{''''})_n^1)
\end{equation}
\begin{equation}
	y = Dropout(BN(ReLU(FFM(x^{'''''}))))
\end{equation}
\noindent
where $CS(\cdot)$ denotes the ConvSubsampling, $P(\cdot)$ refers to the PositionalEncoding, $CB_{stack}(\cdot)$ is the result of stacking multiple ConvBlock, $DA(\cdot)$ represents the DCNN-Attention module, and $DFN_{i+1}^{j}(\cdot)$ means the DualFusionNet as described in the preceding sections. It receives $(2\times j-1)$-th branch and $(2\times j)$-th branch in the $i$-th layer as input, and its output is used as the input of the DCNN-attention module in the $(j)$-th branch in the $(i+1)$-th layer. $SE(\cdot)$ refers to the SENet, it accepts the output of the DCNN-attention in the only branch in the $n$-th layer. $FFM(\cdot)$ signifies the Feed Forward module.

\section{Experiments}
\subsection{\textbf{Dataset}}

We evaluate the proposed model on the Aishell-1 dataset, which consists of 178 hours of labeled speech. The speech sampling frequency is 16kHz and stored in a 16-bit PCM format. The speech is recorded by 400 speakers, which are balanced with their gender, age, and birth-place. The text transcriptions cover 5 domains, i.e., “Finance”, “Science and Technology”, “Sports”, “Entertainments”, and “News”. The dataset is divided into three parts. The training set contains about 150 hours of speech. The development set contains about 18 hours of speech, the test set contains about 10 hours of speech. During training, for convenience, we divide the training set into a training set and development set at a ratio of 0.99 to 0.01, so that we can adjust the model in real-time. The output labels consist of 1304 Chinese characters for Aishell-1.

\subsection{\textbf{Experiments Setup}}
We extracted 80-channel filterbank features computed from a 25ms window with a stride of 10ms. In order to make the model easier to deploy across platforms, we use TensorFlow2\cite{tensorflow2015} to implement the speech spectral feature extraction layer inside the model, the parameters of this layer are about 1.1M. All models are trained on 1 Tesla V100S GPU with batch size 32. For each audio in the same batch, we pad it to the maximum speech length in the batch. Not only that, we use a rate of $P_drop$ = 0.1 for every dropout\cite{2014Dropout} layer. We use the Adam optimizer\cite{2014Adam} with $\beta_{1}$ = 0.9, $\beta_{2}$ = 0.98 and $\epsilon$ = $10^{-6}$ and a warm-up learning rate schedule \cite{2017Attention}, its formula is:
\begin{equation}
	lr = \sqrt{d_{model}} \cdot min(\sqrt{step}, step_{warm-up}^{-1.5})
\end{equation}
\noindent
where $d_{model}$ is the model dimension in multi-branch fusion DCNN network with multi-head self-attention, $step$ is the total number of steps for training, $step_{warm-up}$ is the number of warm-up steps, and we set it to 0.1 times of $step$. Character error rate (CER) are adopted for model evaluation.

We use a transformer language model (LM) trained on the Aishell-1 language model corpus, in which the layer dimension of multi-head self-attention is set to 256, both the encoder and decoder are set to 5 layers. We use encoder and decoder embedding sizes of 512, and 512 respectively. We use 4 headed attention in the self-attention, and its optimizer and learning rate schedule are the same as the acoustic model. In the next subsections, we analyze the effect of our design in the proposed multi-branch fusion DCNN network with multi-head self-attention using the Aishell-1 test sets.

\subsection{\textbf{Results on Aishell-1}}
We select the model architecture with the best performance by combining the number of ConvBlocks, the number of attention heads, the expansion factor of each ConvBlocks, and the number of branches, etc. Table \uppercase\expandafter{\romannumeral1} describes their architecture hyper-parameters. We identify three models, small, medium, and large, with 13.6M, 18.9M, and 28.4M parameters respectively. We set the medium model as the baseline. In the following ablation experiments, the hyperparameter configuration of the model(M) is used as the basic configuration. If it is different from the model(M), we will point it out.

\begin{table}[h]
	\renewcommand{\arraystretch}{1.3}
	\caption{Model hyper-parameters for our proposed S, M, L models, combining the number of ConvBlocks, the number of attention heads, the expansion factor of each ConvBlocks, and the number of branches, the model dimension.}
	\label{table_1}
	\centering
	\begin{tabular}{c|ccc}
		\hline
		Model & Proposed(S) & Proposed(M) & Proposed(L)\\
		\hline
		Num Params(M) & 13.6 & 18.9 & 28.4\\
		\hline
		Num ConvBlock & 8 & 8 & 8\\
		\hline
		Expansion Factor & 2-2-2-2 & 2-2-...-2 & 1-2-2-4-4-2-2-1\\
		\hline
		Attention Heads & 4 & 4 & 8\\
		\hline
		Num Layers & 3 & 4 & 5\\
		\hline
		Num Branches & 4 & 8 & 16\\
		\hline
		Dilated Rate & 1-2-4-8 & 1-2-4-...-14 & 1-2-3-...-16\\
		\hline
		Num Model Dim & 256 & 256 & 256\\
		\hline
	\end{tabular}
\end{table}

\begin{table}[h]
	\renewcommand{\arraystretch}{1.3}
	\caption{CER comparison with previous work in the AISHELL-1 task, The symbol ‘RTF’ represents real-time factor, which is tested on CPU single-core decoding tasks.}
	\label{table_2}
	\centering
	
	\begin{tabular}{cccc}
		\hline
		Model & \#Params (M) & CER (\%) & RTF (s) \\
		\hline
		Low-Rank Transformer\cite{2020Low-Rank-Transformer} & 13 & 13.09 & $-$\\
		Sync-Transformer (6-layer)\cite{2020Sync-Transformer} & $-$ & 8.90 & $-$\\
		KERMIT\cite{2020KERMIT}   & $-$   & 7.80   & $-$\\
		ST-NAR\cite{2020ST-NAR}       & 31    & 7.02   & $-$\\
		HS-DACS\cite{2021HS-DACS} & $-$   & 6.80   & $-$\\
		A-FMLM\cite{2020A-FMLM}   & $-$   & 6.70   & 0.28\\
		Transformer\cite{2018ESPnetTransformer} & 29.7 & 6.70 & $-$\\
		\hline
		\textbf{our Proposed(S)}  & \textbf{13.6}  & 7.20 & 0.15\\
		\textbf{our Proposed(M)}  & 18.9  & 6.91   & 0.23\\
		\textbf{our Proposed(L)}  & 28.4  & \textbf{6.45} & 0.36\\
		\hline
	\end{tabular}
\end{table}

We compared the result(CER) of our model on Aishell-1 test sets with a few models. In\cite{2020Low-Rank-Transformer}, the linear in the multi-head self-attention was replaced by the linear encoder-decoder (LED), which greatly reduces the number of parameters of the transformer. In \cite{2020Sync-Transformer}, a strategy to synchronize codecs in transformer was proposed, which forces each node in the encoder to focus only on its left context, eliminating the dependence of self attention mechanism on future information. The decoder starts to predict symbols immediately after the encoder generates a fixed-length state sequence block. In \cite{2020KERMIT}, a strategy of using CTC loss function in transformer to predict the target sequence length and jointly optimizing with the CE loss function to accelerate convergence was proposed. In \cite{2020ST-NAR}, an insertion based model was proposed to estimate the length of the output token sequence, the CTC was jointly modeled with the insertion-based model. In \cite{2021HS-DACS}, a strategy of synchronizing each attention head in the transformer was proposed to reduce the loss of attention weight.

The results are shown in Table \uppercase\expandafter{\romannumeral2}; all our evaluation results  are rounded to 2 digits after decimal point. Using the language model, our small, medium and large models have achieved excellent CER performance of 7.20\%, 6.91\% and 6.45\%, respectively, in the tests. Compared with the other state-of-the-art systems, it is shown that our design performs best on the test set. Especially for the proposed model(L), when the number of parameters is smaller than that of the best known Transformer\cite{2018ESPnetTransformer}, the results are better than those of the Transformer. In addition, the proposed model (S) has a much smaller number of parameters than other models and achieves a very competitive recognition accuracy. This clearly demonstrates the effectiveness of combining DCNN, multi-head self-attention, and multi-branch progressive fusion network.

\subsection{\textbf{Ablation Studies}}
In this section, the proposed idea of pyramid multi-branch fusion DCNN with multi-head self-attention is verified by ablation tests. For consistency of comparison, all models are trained on 1 Tesla V100S GPU with batch size 32 and training epochs of 90.

\subsubsection{\textbf{Effects of dilated rate}}

We set up multiple experimental groups with the different dilated rate for comparison. The results are shown in Table \uppercase\expandafter{\romannumeral3}. The dilated rate of 8 branches in the first group is all 1, that is, DCNN is not used. Compared with other experimental groups, it can be found that its CER is the highest. In the second set of experiments, we set half of the branches to use DCNN and the other half to disable DCNN, its CER decreased significantly. When we deploy DCNN in each branch, the CER drops again. All these indicate that the DCNN we adopted is effective.

We also tested which combination of dilated rate can achieve lower CER. The dilated rate of the last three experimental groups are arithmetic progressions with differences of 1, 2, and 3, where the combination of dilated rate with a difference of 2 achieved the lowest CER 6.91\%. The third set of results shows that the CER increases when the dilated rate is set too large. This is because the input speech duration is generally not more than 15s, therefore, input speech can only be divided into several hundreds of speech frames, after 4 layers of DCNN extraction with a large dilated rate, these branches all have a global receptive field, they are no longer different. On the other hand, when we set the dilated rate to be less different, such as in the fourth experimental group, the difference in receptive field range of each branch is not significant, so the results are suboptimal. Therefore, we selected the dilated rate combination of the last experimental group.

\begin{table}[h]
	\renewcommand{\arraystretch}{1.3}
	\caption{Effects of different dilated rate on CER(\%). The expansion factor of each ConvBlock is set to 2, that is, the number of convolution kernels of pointwise convolution is $2\times d_{ model}$.}
	\label{table_3}
	\centering
	
	\begin{tabular}{cc}
		\hline
		Dilated Rate & CER (\%) \\
		\hline
		1-1-1-1-1-1-1-1 & 9.02\\
		1-1-1-1-2-3-5-8 & 7.56 \\
		1-3-6-9-12-15-18-21 & 7.23 \\
		1-2-3-4-5-6-7-8 & 7.22 \\
		1-2-4-6-8-10-12-14 & 6.91 \\
		\hline
	\end{tabular}
\end{table}

\subsubsection{\textbf{Effects of multiple branches}}

We tested the effect of different branch numbers on CER and selected three groups with branch numbers of 4, 8, and 16 as comparison experiments. The dilated rate of the experimental group with 4 branches was 1-2-4-6, the dilated rate of the experimental group with 8 branches was 1-2-4-6-8-10-12-14. Regarding the selection of the dilated rate, we know that when the dilated rate is too large, the CER will increase. Therefore, the dilated rate of this experimental group with 16 branches is 1-2-$\cdot\cdot\cdot$-16. The experimental results are shown in Table \uppercase\expandafter{\romannumeral4}. It is clear that CER decreases significantly as the number of branches increases. In particular, the CER of the experimental group with 16 branches decreases by 0.62\% compared with that of the experimental group with 4 branches, but at the same time, the training time increases due to the increase of parameters. Each additional 4 branches will increase about 5M parameters and about 17 hours of training time.

\begin{table}[h]
	\renewcommand{\arraystretch}{1.3}
	\caption{Effects of multiple branches. Training Time shows how long it took the model to train for 90 epochs.}
	\label{table_4}
	\centering
	
	\begin{tabular}{cccc}
		\hline
		Branches Nums & \#Params (M) & CER (\%) & Training Time (hr) \\
		\hline
		4 & 13.6 & 7.20 & $\approx 38$\\
		8 & 18.9 & 6.91 & $\approx 54$ \\
		16 & 28.4 & 6.58 & $\approx 72$ \\
		\hline
	\end{tabular}
\end{table}

\subsubsection{\textbf{Effects of attention head}}

The model dimension is 256, so we conducted experiments when the number of heads is 2, 4, and 8, the dimensions of each head are 128, 64 and 32 respectively, the results are shown in Table \uppercase\expandafter{\romannumeral5}.

It shows that the more the number of heads, the lower the CER of the model. This well answers the questions raised in this paper and also proves the correctness of the combination of DCNN, multi-head self-attention, and multi-branch fusion network. That is when the multi-branch network with different expansion rates can disperse the semantic features to each branch well, it will greatly reduce the features that each branch needs to represent. Therefore, even if the dimension of each head is small, it can fully represent its semantic features.

\begin{table}[h]
	\renewcommand{\arraystretch}{1.3}
	\caption{Effects of attention head. Except for the number of heads, we set the same hyperparameters as proposed(M) for all the following models, 'Dim per Head' denotes the dimension of each attention subspace.}
	\label{table_5}
	\centering
	
	\begin{tabular}{cccc}
		\hline
		Heads Num & Dim per Head & CER (\%) \\
		\hline
		2 & 128 & 8.8 \\
		4 & 64 & 6.91 \\
		8 & 32 & 6.67 \\
		\hline
	\end{tabular}
\end{table}

\subsubsection{\textbf{Effects of SENet, FFModule}}

We have introduced in the third section that SENet weights each channel of the feature matrix through a weight matrix with a global receptive field, so the larger the dimension of the feature matrix, the more detailed the characterization of the features will be by the weighting matrix. In\cite{2020ContextNet}, the dimension of the convolution block also gradually increases, which also proves this. In Table \uppercase\expandafter{\romannumeral6}, we tested CER in four cases: the dimensions of SENet and FFM are 512 and 256 respectively, only SENet, only FFM, and neither.

The experimental results show that both the FFM module and SENet have significantly reduced the CER of the model, by 1.43\%, 1.83\% respectively. Of course, better results were obtained when the two were used together, with a drop of 2.06\%. This undoubtedly proves the effectiveness of these two modules. In addition, we also experimented the effect of the last layer of convolution with a kernel of 256 and 512, respectively. The results were consistent with our assumptions. When the convolution kernel is 256, the representation ability of the feature is far from enough, so only a CER of 10.30\% is obtained. In contrast, with a kernel size of 512, the model achieved a 1.33\% drop in CER. But if we set all the convolution kernel sizes to 512, the model will be very bloated, which is not in line with our original intention, so the strategy of this paper only sets the last layer to 512.

\begin{table}[h]
	\renewcommand{\arraystretch}{1.3}
	\caption{Effects of SENet, FFModule. 'Dim of Last Layer' represents the number of convolution kernels of DCNN in the last layer. 'Dim of SE and FFM' denotes the dimension of SENet, FFModule respectively.}
	\label{table_6}
	\centering
	
	\begin{tabular}{c|ccc}
		\hline
		\multirow{2}{*}{Model} & Dim & Dim  & \multirow{2}{*}{CER (\%)} \\
		& Last Layer & SE and FFM \\
		\hline
		$-$(SENet+FFM)& 512 &(0,0)     & 8.97 \\
		$-$SENet      & 512 &(0,512)   & 7.14 \\
		$-$FFM        & 512 &(512,0)   & 7.54 \\
		SENet+FFM     & 512 &(512,512) & 6.91 \\
		\hline
		$-$(SENet+FFM)& 256 &(0,0)     & 10.30 \\
		SENet+FFM     & 256 &(256,256) & 9.79 \\
		\hline
	\end{tabular}
\end{table}

\subsubsection{\textbf{Effects of ConvBlocks and expansion factor in ConvBlock}}

We tested the effect of ConvBlock 4, 8, 16 on CER of the model, respectively. The results are shown in Table \uppercase\expandafter{\romannumeral7}. We find that the CER of the model decreases with the increase of convolution blocks. When the number of convolution blocks increased from 4 to 8, the CER decreased by 0.2\%. However, when ConvBlock rises to 16, the CER of the model decreases by only 0.05\%, which also indicates that a blind increase does not reduce the CER very well. We also tested the performance of models when ConvBlock's expansion factors are different. Surprisingly, when we used the symmetric structure in the fourth experimental group, the model CER decreased significantly by 0.12\%. This structure has fewer parameters than simply increasing ConvBlocks, so our proposed(L) uses this structure instead of 16 ConvBlocks.

\begin{table}[h]
	\renewcommand{\arraystretch}{1.3}
	\caption{Effects of ConvBlocks and expansion factor in ConvBlock.}
	\label{table_7}
	\centering
	
	\begin{tabular}{cccc}
		\hline
		ConvBlocks & expansion factor & \#Params (M) & CER (\%) \\
		\hline
		4  & 2-2-2-2                   & 17.3  &7.11\\
		8  & 2-2-$\cdot \cdot \cdot$-2 & 18.9  & 6.91 \\
		8  & 1-2-2-2-4-4-4-1           & 20.4  & 6.99 \\
		8  & 1-2-2-4-4-2-2-1           & 19.8  & 6.79 \\
		16 & 2-2-$\cdot \cdot \cdot$-2 & 22.0  & 6.86 \\
		\hline
	\end{tabular}
\end{table}

\section{Conclusion}
In this work, we introduce pyramid multi-branch fusion DCNN with multi-head self-attention, a multi-branch progressive fusion network with a combination of DCNN and attention on each branch. Every two branches fuse until they finally become one. This structure allows us to use a larger number of heads in multi-head self-attention to improve model accuracy. In our ablation experiments, we demonstrate the effectiveness of DCNN, SENet, and other modules respectively. It is verified that the multi-branch fusion mechanism makes our model more competitive with fewer parameters and has obtained 6.45\% CER in the test on of Aishell-1.

\section*{Acknowledgment}
The authors would like to thank the editors and the anonymous reviewers.

\input{article.bbl}


\end{document}